\documentstyle[prl,aps,multicol,epsf]{revtex}
\newcommand{\ket}[1]{| #1 \rangle _{AB}}
\newcommand{\kett}[1]{| #1 \rangle _{A^{\prime}B^{\prime}}}

\begin{document}

\draft

\title{Recovery of entanglement lost in entanglement manipulation}
\author{Fumiaki Morikoshi \cite{E-mail}}
\address{Department of Physics, Hokkaido University, Sapporo 060-0810, Japan}

\maketitle

\begin{abstract}

When an entangled state is transformed into another one with probability one
by local operations and classical communication,
the quantity of entanglement decreases.
This letter shows that entanglement lost in the manipulation can be
partially recovered by an auxiliary entangled pair.
As an application, a maximally entangled pair can be obtained
from two partially entangled pairs with probability one.
Finally, this recovery scheme reveals a fundamental property of entanglement
relevant to the existence of incomparable states.

\end{abstract}


\begin{multicols}{2}
\narrowtext

Quantum entanglement plays an important role in quantum information processing.
It realizes novel information processing that is impossible
in a classical manner.
Thus, in addition to practical applications,
quantum entanglement itself has been widely studied in recent years.
For a detailed review, see Ref.\ \cite{Plenio} and references therein.

One of the most fundamental applications of an entangled state
is quantum teleportation \cite{Bennett93}.
In teleportation, Alice sends a qubit to Bob via a previously shared
maximally entangled state between them,
\begin{equation}
\ket{\Phi^{+}} = \frac{1}{\sqrt{2}} (\ket{00} + \ket{11}).
\label{eq:Bell}
\end{equation}
We refer to the state $\ket{\Phi^{+}}$ as a Bell pair in the following.
All the operations needed are local operations on their respective systems
and classical communication between them.
Since Alice and Bob are distantly located,
they cannot jointly perform global operations on the composite system.
This is always the case in all applications of entangled states
such as quantum communication and quantum cryptography.
Therefore the following question is crucial to
understanding the nature of entangled states.
What can we do on entangled states
by local operations and classical communication alone?

Recently Nielsen found necessary and sufficient conditions
for an entangled state to be transformed into another one
by local operations and classical communication \cite{Nielsen}.
It was also proved that the quantity of entanglement decreases
during the transformation.

It is natural to wish that entanglement would not decrease
because it is a valuable resource.
This letter shows that entanglement lost in entanglement manipulation
can be partially recovered by an auxiliary entangled pair.
Besides the original entangled state to be transformed,
we prepare another entangled state and perform collective operations
on these two pairs.
This transformation enables a part of entanglement lost in the original pair
to be transferred to the auxiliary one.
Entanglement of the whole system decreases during the transformation
in this case too, as required by Nielsen's result.
But this scheme realizes the partial recovery of entanglement 
that is absolutely impossible by individual manipulations of each pair.

As a particular example of the recovery procedure,
it is also shown that we can obtain a Bell pair with probability one
from two partially entangled pairs satisfying a certain condition.

Furthermore, the condition for this recovery scheme to work
reveals a fundamental property of entanglement.
The property has a deep connection with the fact that
there exist essentially different types of bipartite pure--state entanglement,
namely, incomparable states \cite{Nielsen}.

In this letter, we will obtain the main result using Nielsen's theorem.
First, we introduce a mathematical notion of majorization
that is needed in the theorem and is also a main tool in this letter.
Let $x=(x_1, \cdots, x_n)$ and $y=(y_1, \cdots, y_n)$ be
real n-dimensional vectors.
Let $x^{\downarrow}=(x^{\downarrow}_1, \cdots, x^{\downarrow}_n)$ be
the vector obtained by rearranging the elements of $x$ in the decreasing order,
i.e., $x^{\downarrow}_1 \geq \cdots \geq x^{\downarrow}_n$.
We say that $x$ is majorized by $y$, written in $x \prec y$, if
\begin{equation}
 \sum_{j=1}^{k} x^{\downarrow}_{j} \leq \sum_{j=1}^{k} y^{\downarrow}_{j},
 \qquad 1 \leq k \leq n-1,
\label{eq:majorization}
\end{equation}
and
\begin{equation}
 \sum_{j=1}^{n} x^{\downarrow}_{j} =
 \sum_{j=1}^{n} y^{\downarrow}_{j}.
\label{eq:equality} 
\end{equation}

This letter deals with only bipartite pure entangled states,
which are described in Schmidt decomposition such as
$\ket{\psi} = \sum _{i} \sqrt{a_{i}} |i\rangle_{A} |i\rangle_{B}$
where $\{a_{i}\}$ are positive real numbers satisfying
the normalization condition $\sum_{i} a_{i} = 1$.
In Schmidt decomposition,
$\{|i\rangle_{A}\}$ and $\{|i\rangle_{B} \}$ are orthonormal basis of
respective systems; thus eigenvalues of the reduced density matrix
$\rho_{\psi} \equiv tr_{B}(|\psi \rangle _{AB \, AB} \langle \psi|)$ are
$a_{1}, \cdots, a_{n}$.
We define the vector of these eigenvalues as
$\lambda_{\psi} \equiv (a_{1}, \cdots, a_{n})$.
With the theory of majorization,
Nielsen proved the following theorem \cite{Nielsen}.

{\bf Theorem}: A bipartite pure entangled state $\ket{\psi}$
is transformed into another one $\ket{\phi}$
with probability one by local operations and classical communication
if and only if $\lambda_{\psi}$ is majorized by $\lambda_{\phi}$, i.e.,
\begin{equation}
 \ket{\psi} \to \ket{\phi} \quad \text{iff}
  \quad \lambda_{\psi} \prec \lambda_{\phi}.
\label{eq:Nielsen}
\end{equation}

It was also proved that the quantity of entanglement $E(\psi)$,
which is uniquely defined as the von Neumann entropy of $\rho_{\psi}$
\cite{Bennett96,Popescu},
decreases during the transformation,
\begin{equation}
 \text{if} \quad \ket{\psi} \to \ket{\phi}, \quad
 \text{then} \quad E(\psi) \geq E(\phi).
\label{eq:decrease} 
\end{equation}
This follows from the mathematical theorem that
if $ \lambda_{\psi} \prec \lambda_{\phi}$, then $E(\psi) \geq E(\phi)$,
together with Eq.\ (\ref{eq:Nielsen}).
Equation (\ref{eq:decrease}) means that local operations and
classical communication always reduce entanglement.

However, we want to prevent entanglement from decreasing as far as possible,
since we have to send qubits without teleportation
in order to share entanglement between distant observers again.
We will show that an auxiliary entangled pair can partially recover
the entanglement lost in the manipulation of two--qubit entangled states.

The recovery scheme presented in this letter goes as follows.
Suppose we originally want to transform $\ket{\psi}$ into $\ket{\phi}$.
(In the following, we exclude the trivial case $\ket{\psi} = \ket{\phi}$.)
We prepare another entangled state $\kett{\omega}$ besides the system $AB$.
Then we perform collective operations on $\ket{\psi} \otimes \kett{\omega}$,
and convert it to $\ket{\phi} \otimes \kett{\chi}$
where $\kett{\chi}$ has more entanglement than $\kett{\omega}$.
This transformation transfers a part of the entanglement
lost in the system $AB$ to the system $A'B'$.

In the following, it is proved that this scheme is really possible.
We begin with a concrete example to understand the idea of this scheme,
then proceed to a general proof.
We deal with the following example:
\begin{equation}
\begin{array}{rcl}
 \ket{\psi} &=& \sqrt{0.7} \, \ket{00} + \sqrt{0.3} \, \ket{11}, \\
 \ket{\phi} &=& \sqrt{0.8} \, \ket{00} + \sqrt{0.2} \, \ket{11}, \\
 \kett{\omega} &=& \sqrt{0.6} \, \kett{00} + \sqrt{0.4} \, \kett{11}, \\
 \kett{\chi} &=& \sqrt{0.55} \, \kett{00} + \sqrt{0.45} \, \kett{11}.
\end{array}
\end{equation}
The vectors of eigenvalues are
\begin{equation}
\begin{array}{rcl}
 \lambda_{\psi} &=& (0.7, \, 0.3), \\
 \lambda_{\phi} &=& (0.8, \, 0.2), \\
 \lambda_{\omega} &=& (0.6, \, 0.4), \\
 \lambda_{\chi} &=& (0.55, \, 0.45).
\end{array}
\end{equation}
Majorization relations $\lambda_{\psi} \prec \lambda_{\phi}$ and
$\lambda_{\omega} \succ \lambda_{\chi}$ indicate
\begin{equation}
 \ket{\psi} \to \ket{\phi}, \quad \kett{\omega} \gets \kett{\chi},
\end{equation}
and
\begin{equation} 
 E(\psi) \geq E(\phi), \quad E(\omega) \leq E(\chi).
\label{eq:recover}
\end{equation}
If we consider two entangled pairs as one system,
the whole system is an entangled state with Schmidt number four, thus
\begin{eqnarray}
 \lambda_{\psi \otimes \omega} &=& \lambda_{\psi} \otimes \lambda_{\omega} =
 (0.42, \, 0.28, \, 0.18, \, 0.12), \\
 \lambda_{\phi \otimes \chi} &=& \lambda_{\phi} \otimes \lambda_{\chi} =
 (0.44, \, 0.36, \, 0.11, \, 0.09).
\end{eqnarray}
According to Eq.\ (\ref{eq:majorization}), three inequalities
$0.42<0.44$, $0.42+0.28<0.44+0.36$, and $1-0.12<1-0.09$
show that $\lambda_{\psi \otimes \omega} \prec \lambda_{\phi \otimes \chi}$.
[The equality (\ref{eq:equality}) is satisfied by normalization conditions.]
Therefore we can transform $\ket{\psi} \otimes \kett{\omega}$ into
$\ket{\phi} \otimes \kett{\chi}$ by collective manipulation
according to Nielsen's theorem.
Equation (\ref{eq:recover}) means that entanglement lost in the system $AB$ is
partially recovered by the system $A'B'$.
The system $AB$ has no difference between the collective manipulation
and the individual one.
As for the system $A'B'$,
this collective method realizes increase in entanglement,
which cannot be done individually.

Next we prove that the recovery as stated above is always possible.
We find the condition where there exist auxiliary states
$\kett{\omega}$ and $\kett{\chi}$ such that
$E(\omega) < E(\chi)$ and
$\lambda_{\psi \otimes \omega} \prec \lambda_{\phi \otimes \chi}$,
provided that $\ket{\psi} \to \ket{\phi}$.
Let
\begin{equation}
\begin{array}{rcl}
 \ket{\psi} &=& \sqrt{a} \ \ket{00} + \sqrt{1-a} \ \ket{11}, \\
 \ket{\phi} &=& \sqrt{b} \ \ket{00} + \sqrt{1-b} \ \ket{11}, \\
 \kett{\omega} &=& \sqrt{p} \ \kett{00} + \sqrt{1-p} \ \kett{11}, \\
 \kett{\chi} &=& \sqrt{q} \ \kett{00} + \sqrt{1-q} \ \kett{11}.
\end{array}
\end{equation}
The assumption $\ket{\psi} \to \ket{\phi}$ gives
\begin{equation}
 \frac{1}{2} \leq a < b \leq 1.
\label{eq:ab}
\end{equation}
The condition $E(\omega) < E(\chi)$ requires
\begin{equation}
 \frac{1}{2} \leq q < p \leq 1,
\label{eq:pq}
\end{equation}
because $E(\omega) \leq E(\chi)$ is equivalent to
$\lambda_{\omega} \prec \lambda_{\chi}$
in the case of two--qubit states
and the equality holds only for $p=q$.
Combining the two pairs $AB$ and $A'B'$, we have
\begin{eqnarray}
 \lambda_{\psi \otimes \omega} &=& (ap, \, a(1-p), \, (1-a)p, \, (1-a)(1-p)),
\label{eq:ap} \\ 
 \lambda_{\phi \otimes \chi} &=& (bq, \, b(1-q), \, (1-b)q, \, (1-b)(1-q)).
\label{eq:bq}
\end{eqnarray}

In the following, we seek a pair of numbers $(p, q)$
that satisfies the majorization condition
$\lambda_{\psi \otimes \omega} \prec \lambda_{\phi \otimes \chi}$
and Eq.\ (\ref{eq:pq}) with the assumption (\ref{eq:ab}).

The majorization relation
$\lambda_{\psi \otimes \omega} \prec \lambda_{\phi \otimes \chi}$
consists of three inequalities.
[The equality (\ref{eq:equality}) is satisfied
 by normalization conditions.]
We have to rearrange the components of the vectors
(\ref{eq:ap}) and (\ref{eq:bq}) in the decreasing order
before imposing the inequality conditions (\ref{eq:majorization}).
Equations (\ref{eq:ab}) and (\ref{eq:pq}) indicate that
the largest and the smallest elements in (\ref{eq:ap}) are
$ap$ and $(1-a)(1-p)$, respectively. 
Similarly, $bq$ and $(1-b)(1-q)$ are the largest and the smallest
elements in (\ref{eq:bq}), respectively.
Thus the first and the third inequalities of the majorization condition are
$ap \leq bq$ and $1-(1-a)(1-p) \leq 1-(1-b)(1-q)$, i.e.,
\begin{eqnarray}
 q &\geq& \frac{a}{b}\,p,
\label{eq:dashdotted} \\
 1-q &\leq& \frac{1-a}{1-b} (1-p), 
\label{eq:dashed}
\end{eqnarray}
where Eq.\ (\ref{eq:ab}) implies
\begin{equation}
 \frac{1}{2} \leq \frac{a}{b} < 1, \quad 1 < \frac{1-a}{1-b}.
\label{eq:range}
\end{equation}

However, Eqs.\ (\ref{eq:ab}) and (\ref{eq:pq}) cannot tell
which is the next largest element in (\ref{eq:ap}) and (\ref{eq:bq}).
If $a \leq p$ then $a(1-p) \leq p(1-a)$, and so on.
Thus comparing the second and the third elements in each vector,
we have the following three cases:
(i) $a \leq p,\: b \leq q$, (ii) $ a \leq p,\, b >q$, (iii) $a>p,\, b>q$.
[The case $a>p, \, b \leq q$ contradicts
 Eqs.\ (\ref{eq:ab}) and (\ref{eq:pq}).]

(i) $a \leq p, \, b \leq q$: \
The next largest elements in (\ref{eq:ap}) and (\ref{eq:bq}) are
$(1-a)p$ and $(1-b)q$, respectively.
Thus the second inequality of the majorization condition is
$ap+(1-a)p \leq bq+(1-b)q$, \,i.e., \, $p \leq q$,
which contradicts Eq.\ (\ref{eq:pq}).

(ii) $a \leq p, \, b>q$: \
Since the elements $(1-a)p$ and $b(1-q)$ are the next largest elements
in (\ref{eq:ap}) and (\ref{eq:bq}), respectively,
we have $ap+(1-a)p \leq bq+b(1-q)$, i.e., $p \leq b$.
Therefore,
\begin{equation}
 a \leq p \leq b, \quad b>q.
\label{eq:case2}
\end{equation}

(iii) $a>p, \, b>q$: \
Similarly, the majorization condition requires
$a \leq b$, which is implied in Eq.\ (\ref{eq:ab}).
In this region, we have
\begin{equation}
 a>p, \quad b>q.
\label{eq:case3}
\end{equation}

Summing up these cases, we see that
the second inequality of the majorization condition is
Eq.\ (\ref{eq:case2}) or (\ref{eq:case3}):
\begin{equation}
 p \leq b, \quad q < b
\label{eq:maj2}\end{equation}

\begin{figure}
\begin{center}
\begin{minipage}{6.5cm}
\epsfxsize=6.5cm \epsfbox {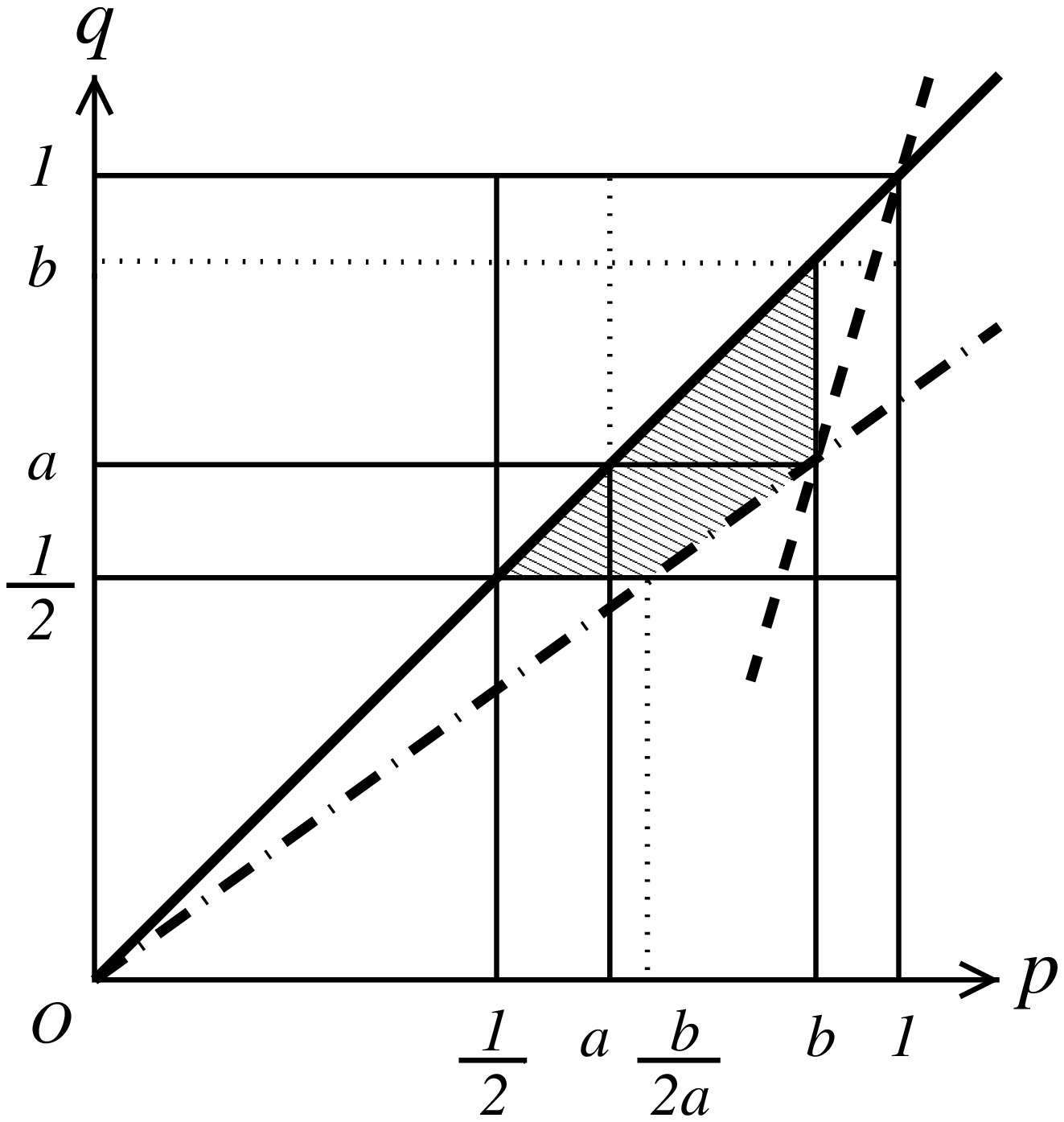}
\end{minipage}
\end{center}
\caption{The condition for the recovery scheme is satisfied in the shaded
 region. The thick solid line, the broken line, and the dash-dotted line
 represent $q=p$, $1-q=\{(1-a)/(1-b)\}(1-p)$, and $q=(a/b)p$, respectively.}
\label{Fig1}
\end{figure}

As a result, $(p,q)$ must satisfy Eqs.\ (\ref{eq:pq}),
(\ref{eq:dashdotted}) -- (\ref{eq:range}), and (\ref{eq:maj2}).
Figure \ref{Fig1} shows these conditions as a shaded region in $p-q$ plane.
It indicates that there exists the shaded region
irrespective of $a$ and $b$.
Thus if we take the auxiliary states $\kett{\omega}$ and $\kett{\chi}$
appropriately, recovery of entanglement is always possible.

Now, we discuss the implication of Fig.\ \ref{Fig1} in detail.
The shaded region in Fig.\ \ref{Fig1} is divided into two parts,
$q\geq a$ and $q<a$.
In the region $q \geq a$, we have
$\frac{1}{2} \leq a \leq q < p \leq b \leq 1$.
This means that $\lambda_{\psi} \prec \lambda_{\chi}$ and
$\lambda_{\omega} \prec \lambda_{\phi}$.
If we perform  $\ket{\psi} \to \ket{\chi}, \kett{\omega} \to \kett{\phi}$,
and interchange $AB$ and $A'B'$,
then the recovery process stated above is also accomplished
by only the individual manipulations of each pair.
Therefore this region of Fig.\ \ref{Fig1} presents
trivial recovery that needs no collective manipulation.
However, in the region $q<a$, we have
$ \frac{1}{2} \leq q < a \leq p \leq b \leq 1 $, or
$ \frac{1}{2} \leq q < p < a < b \leq 1 $.
These inequalities imply that neither $| \psi \rangle$ nor $|\omega \rangle$
can be transformed into $|\chi \rangle$.
Thus this region presents true recovery that only the collective
manipulation can realize.
In fact, we do not need the trivial region for recovery,
because, for each point in the trivial region,
there exist points with the same $p$ and
the smaller $q$ in the true region.
Only the true region, the shaded part in $q<a$, is of great importance.

It should also be noted that the complete recovery is represented
only at the point $(b,a)$ in Fig.\ \ref{Fig1},
which corresponds to the trivial interchange of $AB$ and $A'B'$.

A useful application of this scheme is to obtain a Bell pair
[Eq.\ (\ref{eq:Bell})] after a recovery procedure.
Figure \ref{Fig1} shows that if we prepare $\kett{\omega}$ having $p$ such that
$p \leq b/(2a)$,
we can transform the $A'B'$ pair into a Bell pair with probability one.
In addition to the Bell pair,
there exists residual entanglement in the system $AB$.
If we does not need this residual entanglement in $AB$, which means $b=1$,
a Bell pair can always be obtained from two partially entangled pairs
$\ket{\psi}, \kett{\omega}$ such that
\begin{equation}
 ap < \frac{1}{2}.
\label{eq:concentration}
\end{equation}
An explicit example of this concentration is as follows:
\begin{equation}
\begin{array}{rcl}
 \ket{\psi} &=& \sqrt{0.6}  \ \ket{00} + \sqrt{0.4} \ \ket{11}, \\
 \ket{\phi} &=& \sqrt{0.9} \ \ket{00} + \sqrt{0.1} \ \ket{11}, \\
 \kett{\omega} &=& \sqrt{0.7} \ \kett{00} + \sqrt{0.3} \ \kett{11}, \\
 \kett{\chi} &=& \sqrt{0.5} \ \kett{00} + \sqrt{0.5} \ \kett{11}\\
             &=& \kett{\Phi^{+}}.
\end{array}
\end{equation}
The eigenvalues of the product states are
\begin{eqnarray}
 \lambda_{\psi \otimes \omega} &=& \lambda_{\psi} \otimes \lambda_{\omega} =
 (0.42, \, 0.18, \, 0.28, \, 0.12), \\
 \lambda_{\phi \otimes \chi} &=& \lambda_{\phi} \otimes \lambda_{\chi} =
 (0.45, \, 0.45, \, 0.05, \, 0.05).
\end{eqnarray}
Thus $\lambda_{\psi \otimes \omega} \prec \lambda_{\phi \otimes \chi}$
indicates that the concentration
$\ket{\psi} \otimes \kett{\omega} \to \ket{\phi} \otimes \kett{\Phi^{+}}$
is possible with probability one.

The Procrustean method \cite{Bennett96} is already known as a way of
obtaining a Bell pair from a partially entangled state.
Since this method works only probabilistically, however,
we cannot necessarily obtain a Bell pair by applying
the method to partially entangled pairs.
Thus this application of the recovery scheme is very important
for practical purpose.
If there happen to be two partially entangled pairs satisfying
Eq.\ (\ref{eq:concentration}),
then we can always prepare a Bell pair from them for future use.

The collective manipulation of both pairs is the heart of this recovery scheme.
It makes the transformation possible that is absolutely impossible
by individual manipulations of each pair.
This is reminiscent of the reversibility between entanglement concentration
and dilution in the asymptotic limit \cite{Bennett96,Popescu}
and the catalysis in entanglement manipulation discovered in \cite{Jonathan}. 

\begin{figure}
\begin{center}
\begin{minipage}{6.5cm}
\epsfxsize=6.5cm \epsfbox {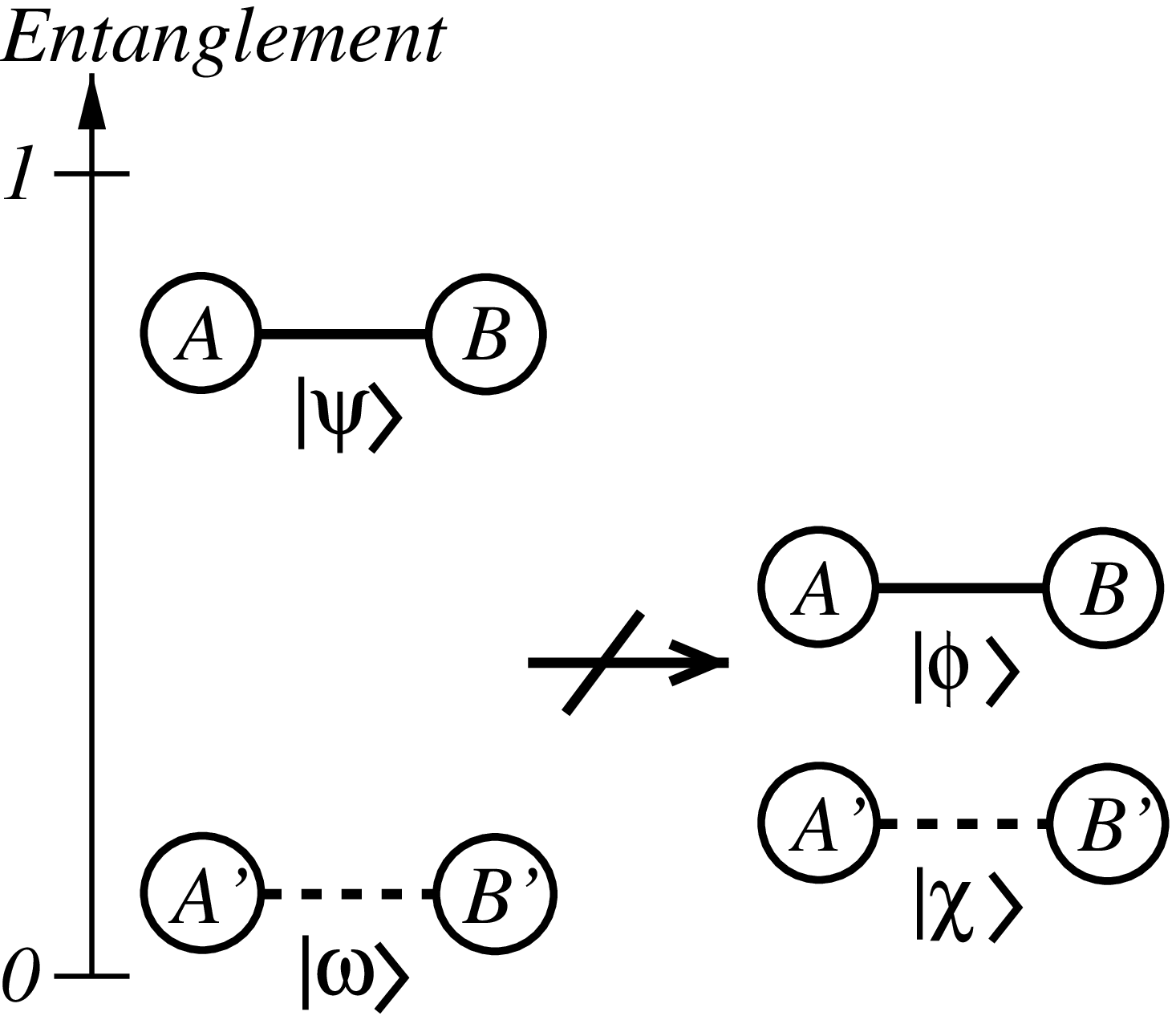}
\end{minipage}
\end{center}
\caption{Each pair connected by a solid or dashed line represents
 an entangled state. The axis indicates the quantity of entanglement.
 If $\kett{\omega}$ has less entanglement than $\ket{\phi}$,
 the recovery process is impossible no matter how much entanglement
 is lost in the system $AB$.}
\label{Fig2}
\end{figure}

Finally we consider a fundamental property of entanglement
that this recovery scheme reveals.
The most striking part of the condition described in Fig.\ \ref{Fig1}
is $p \leq b$.
This condition implies that, if we intend to recover the entanglement
lost in the transformation $\ket{\psi} \to \ket{\phi}$,
we must prepare the auxiliary state $\kett{\omega}$
that has more entanglement than $\ket{\phi}$.
It depends on only the final state $\ket{\phi}$,
not the quantity of entanglement lost in the transformation
$\ket{\psi} \to \ket{\phi}$,
whether the recovery procedure by $\kett{\omega}$ is possible or not.
No matter how much entanglement is lost,
nothing can be recovered if the auxiliary state is not sufficiently entangled.
This is the fundamental property of bipartite pure entangled states
revealed by the recovery scheme.
This surprising feature of entanglement is depicted in Fig.\ \ref{Fig2}.
The notion of entanglement measure cannot fully explain this property.

This new property of entanglement is a direct consequence of 
the existence of incomparable states \cite{Nielsen}.
The states $\ket{\alpha}$ and $\ket{\beta}$ are called incomparable
if neither $\ket{\alpha} \to \ket{\beta}$ nor $\ket{\beta} \to \ket{\alpha}$.
If $p$ is greater than $b$, then the second inequality of
the majorization condition 
$\lambda_{\psi \otimes \omega} \prec \lambda_{\phi \otimes \chi}$
is not satisfied.
Taking into account other inequalities of the majorization condition,
we see that
$\ket{\psi} \otimes \kett{\omega}$ and $\ket{\phi} \otimes \kett{\chi}$
are incomparable in the region $b<p\leq1, \, (a/b)p\leq q<p$.
[In the region $b<p \leq 1, \, (1/2) \leq q <(a/b)p$,
 the entanglement of the whole system increases because of
 $\lambda_{\psi \otimes \omega} \succ \lambda_{\phi \otimes \chi}$.
 Thus this region is excluded by Nielsen's result.]
Therefore the impossibility of recovery by an insufficiently
entangled pair is directly connected to the existence of incomparable states.

In conclusion,
we have proved that entanglement lost in entanglement manipulation
can be partially recovered by an auxiliary entangled pair.
This recovery scheme has also revealed the fundamental property of
quantum entanglement that has a connection with
the existence of incomparable states:
When we intend to transfer entanglement from one pair to another,
nothing can be transferred if the recipient is not sufficiently entangled.
More detailed investigations are necessary to grasp the deep implication
of this property.

I am grateful to K. Suehiro for a careful reading of the manuscript.

\end{multicols}

\end{document}